# Metallo-Anti-aromatic Al$_4$Na$_4$ and Al$_4$Na$_3^-$ compounds: A Theoretical Investigation.

Sharan Shetty, Dilip G. Kanhere, Sourav Pal[*]

[*]  Sourav Pal, Sharan Shetty

Theoretical Chemistry Group,

Physical Chemistry Division,

National Chemical Laboratory,

Pune-411008, India

Fax:- +91-20-5893044

E-mail:- **pal@ems.ncl.res.in**

Dilip G. Kanhere

Department of Physics,

University of Pune,

Pune-411007, India

**Summary**

Present theoretical investigation reveals that the neutral Al$_4$Na$_4$ and anion Al$_4$Na$_3^-$ clusters satisfy the criteria for anti-aromaticity and are metallo-anti-aromatic compounds.

It has been understood for many years that amongst the simple monocyclic systems, compounds with only (4n+2) π-electrons (where n is an integer), such as benzene, follow the Hückel's rule and are aromatic.[1] Conjugated ring systems having (4n) π-electrons do not obey Hückel's rule for aromaticity and can be divided into anti-aromatic and non-aromatic compounds. The definition of anti-aromaticity is somewhat controversial, but the concept has proven to be of great interest due to the unusual behavior exhibited by the anti-aromatic compounds, like instability, magnetic properties and high reactivity.[2] If we restrict to the conventional definition of anti-aromaticity, then it is known that if the system is planar with (4n) π-electrons and is destabilized due to the electron delocalization, it will be anti-aromatic e.g. cyclobutadiene.[3,4] On the other hand if the system with (4n) π-electrons buckle to become non-planar, it is known to loose the anti-aromaticity and the system would be non-aromatic, e.g. cyclooctatetraene.[5] However, the distinction between anti-aromaticity and non-aromaticity is not very clear. The geometry of the system seems to be one criterion to distinguish anti-aromatic and non-aromatic compounds. Cyclobutadiene is a well-known example of anti-aromatic compounds.[3,4]

Aromaticity and anti-aromaticity have been historically important in organic chemistry. However, in recent years, studies have been carried out to show that aromaticity also exists in organometallic compounds and metal clusters.[6,7,8] In a combined experimental and theoretical work, Li *et al* have shown aromaticity in all-metal atom clusters e.g. $MAl_4^-$ and $M_2Al_4$ (M = Li, Na and Cu) for the first time.[8(a)] On this basis, recently it was shown that $Al_4^{4-}$ species in $Al_4Li_4$ cluster satisfies the criteria of anti-aromaticity and hence is an all-metal anti-aromatic compound.[9,10] Theoretical proposition of this was presented by us in a recent symposium.[9] However, Kuznetsov *et al* have experimentally synthesized the first anti-aromatic compound viz. $Al_4Li_3^-$ and theoretically proved that it also has similar characteristics of $Al_4Li_4$ cluster in a receent publication.[10] These studies opens up a new area in the field of aromaticity and anti-aromaticty in metal compounds.

On this background, we propose a theoretical investigation in this paper to understand the structural and bonding properties in neutral $Al_4Na_4$ and anion $Al_4Na_3^-$ clusters. We show that the $Al_4$ species in neutral $Al_4Na_4$ and anion $Al_4Na_3^-$ is similar to the $Al_4$ species found in $Al_4Li_4$ and $Al_4Li_3^-$ clusters and are anti-aromatic.

Before we discuss the results on bonding, we present the ground state geometry of the $Al_4Na_4$ and $Al_4Na_3^-$ clusters. The structural property of $Al_4Na_4$ cluster has been already studied using Born-Oppenhiemer molecular dynamics (BOMD) simulations.[11] The ground state geometries obtained through BOMD simulations were used as starting geometries for *ab initio* calculations.[12] We found that the ground state geometry obtained from the *ab initio* calculations for $Al_4Na_4$ (Fig. 1(a)) is a capped octahedron ($C_{2v}$ symmetry). The four Al atoms form a rectangular planar structure ($D_{2h}$ symmetry) with two of its bond lengths being 2.68 Å and the other two being 2.85 Å. The rectangular planar $Al_4$ structure found in the $Al_4Na_4$ cluster is similar to the $Al_4$ structure found in the $Al_4Li_4$ cluster studied earlier.[9,10] The excited state geometry of $Al_4Na_4$ (Fig. 1(b)) is seen to form a quinted roof with a buckled $Al_4$ unit and is 0.3-0.4 kcal/mol higher in energy than the ground state. However, the ground state $Al_4Na_4$ structure obtained through BOMD calculations showed that the $Al_4$ species is buckled (Fig. 1(b)).[11] The ground state geometry of $Al_4Na_3^-$ cluster (Fig. 1(c)) is also a capped octahedron ($C_{2v}$ symmetry). The

$Al_4Na_3^-$ system, having rectangular planar $Al_4$ structure ($D_{2h}$ symmetry), is similar to the $Al_4$ structure found in the $Al_4Na_4$ cluster, with the only difference that the shorter Al-Al bond is 2.28 Å and longer Al-Al bond is 2.44 Å unlike the slightly distorted rectangular $Al_4$ structure in $Al_4Li_3^-$ cluster.[10] This is due to the symmetric capping of the three Na atoms. It is also seen that the $Al_4$ unit gets contracted in the anion then in the neutral species. We have also carried out spin polarized calculations for $Al_4Na_4$ and $Al_4Na_3^-$ system and it was found that the singlet state is more stable than the triplet state.

The hybridization of the four Al atoms in the $Al_4Na_4$ and $Al_4Na_3^-$ clusters may be considered as $sp^2$, leaving one empty un-hybridized p-orbital on each Al atom in the $Al_4$ species. Although the difference in the ionization potential of Al and the Na atoms is very small, earlier studies have shown that in these classes of hetero-clusters, the more electronegative atom occupies the interstitial position and behaves as a single entity or superatom.[11, 15] In the present study, the $Al_4$ species in $Al_4Na_4$ and $Al_4Na_3^-$ clusters behaves as a superatom and hence the electron affinity of the $Al_4$ species increases compared to the alkali atoms. This arrangement of the $Al_4$ species in $Al_4Na_4$ and $Al_4Na_3^-$ clusters allows the Na atoms to donate one electron to the unoccupied $p_z$-orbital of the four Al atoms, thus providing the required 4π electrons for anti-aromaticity. The highest occupied molecular orbital (HOMO) picture (Fig. 2) clearly shows two localized π-bonds along the two Al atoms having shorter bond lengths (2.68 Å) of the $Al_4$ unit in $Al_4Na_4$ cluster. HOMO-1 (Fig. 2) is mainly a transannular bonding between the pair of σ bonds of Al-Al atoms and lone pair of electrons of π character is seen. This kind of transannular bonding was also seen in the $Ga_4H_2^{2-}$ compound.[16] HOMO-2 (Fig. 2) also shows a similar kind of transannular bonding between the pair of Al-Al bonds having higher bond lengths. HOMO-3 (Fig. 2) also shows a transannular bonding, but it is seen between the π orbitals of the pair of Al-Al bonds. HOMO-4 (Fig. 2) shows lone pair of electrons on the four Al atoms. The two localized π-bonds in HOMO of $Al_4Na_4$ cluster (Fig. 2) arise due to the 4π electrons donated by the four Na atoms to the $Al_4$ species, which is consistent with the charge transfer from the Na atoms to the Al atoms.

Interestingly, the bonding nature in $Al_4Na_3^-$ cluster is similar to that in $Al_4Na_4$ cluster. The HOMO (Fig. 3) of $Al_4Na_3^-$ is same as the HOMO (Fig. 2) of $Al_4Na_4$ cluster with alternate π bonds along the shorter Al-Al (2.28 Å) bonds. HOMO-1 (Fig. 3) has a transannular bonding of σ-bond character with lone pair of electrons similar to HOMO-1 of $Al_4Na_4$ system. HOMO-2 (Fig. 3) of $Al_4Na_3^-$ has a transannular bonding between the two Al-Al bonds and also π type lone pair of electrons showing a strong resemblance to the HOMO-2 (Fig. 2) of $Al_4Na_4$. HOMO-3 (Fig. 3) shows localized π-bonds along the shorter Al-Al bonds in the $Al_4$ species but surprisingly a delocalized orbital is also seen within the four Al atoms.

We have also performed the above calculations on $Al_4^{4-}$ system (not shown). Surprisingly, the calculations show that $Al_4^{4-}$ system was highly unstable. We do agree with the fact that, like the instability of $Al_4^{2-}$ system discussed earlier,[16] the instability in the $Al_4^{4-}$ system is due to the coulombic repulsion from the four negative charges on $Al_4^{4-}$. This indicates that the presence of the cation $Na^{4+}$ and $Na^{3+}$ in the $Al_4Na_4$ and $Al_4Na_3^-$ cluster respectively are required for the stability of the $Al_4$ species.

The preference of the rectangular geometry of $Al_4$ species in $Al_4Na_4$ and $Al_4Na_3^-$ cluster, which is a singlet state, as discussed earlier, is due to the mixing of energetically close states (pseudo Jahn-Teller or second order Jahn-teller effect).[4(a),17] This can also be explained by the fact that the distortion of the π electrons drives the molecule to bond alternated geometry where short and localized π bonds can be achieved.[18]

Our investigation shows that the $Al_4$ species in the $Al_4Na_4$ and $Al_4Na_3^-$ system has 4π electrons with a planar rectangular geometry (singlet state) and two localized π bonds. Analogous to the $Al_4Li_4$ and $Al_4Li_3^-$,[10] the present discussion demonstrates that the neutral $Al_4Na_4$ and anion $Al_4Na_3^-$ clusters are metal cyclodienes and are strong candidates for metallo-anti-aromaticity. Thus, the present investigation and the earlier studies[10] show that the anti-aromatic behavior is not only confined to the organic compounds, but is also possible in the metal compounds. These studies on the existence of metallo-antiaromatic compounds would motivate the experimentalists to understand the chemical properties such as stability, reactivity and magnetic properties and to compare them with the anti-aromatic organic compounds.

**ACKNOWLEDGEMENT**
S. Shetty and S. Pal gratefully acknowledge the Indo-French Center for the Promotion of Advance Research (IFCPAR) (Project. No. 2605-2), New Delhi, India, for financial assistance. D. G. Kanhere also acknowledges the grant from the IFCPAR (Project. No 1901), NewDelhi,India.

**Figure Captions**

**Figure. 1.**

Geometries of $Al_4Na_4$ and $Al_4Na_3^-$ clusters (at the CCD/6-31G(d,p) level), black spheres indicate the Al atoms while white spheres indicate the Na atoms. (a) Ground state structure of $Al_4Na_4$ cluster with a reectangular planar $Al_4$ unit (b) Excited state structure of $Al_4Na_4$ cluster with a distorted $Al_4$ unit. (c) Ground state structure of $Al_4Na_3^-$ cluster with a rectangular planar $Al_4$ unit

**Figure. 2**

The last five highest occupied molecular orbital (HOMO) pictures of $Al_4Na_4$ cluster are shown.

HOMO shows a localized and alternate π-orbitals in $Al_4$ plane. HOMO-1 is a transannular bonding between the two pairs of σ-bonds Al-Al and also lone pair of electrons. HOMO-2 is a transannular bonding between the pair of σ-orbitals of Al-Al. HOMO-3 is a transannualr bonding between the two pi-orbitals of the Al-Al bonds. HOMO-4 shows lone pair of electrons on 4 Al atoms.

**Figure. 3**

The last five highest occupied molecular orbital (HOMO) pictures of $Al_4Na_3^-$ cluster are shown.

HOMO shows a localized and alternate π-orbitals in $Al_4$ plane. HOMO-1 is a transannular bonding between the two pairs of σ-bonds of Al-Al. HOMO-2 is a transannular bonding between the two Al-Al bonds and also lone pair of electrons. HOMO-3 is a transannular bonding of π character. HOMO-4 shows localied π-bonds on Al-Al and a delocalized π electron density.

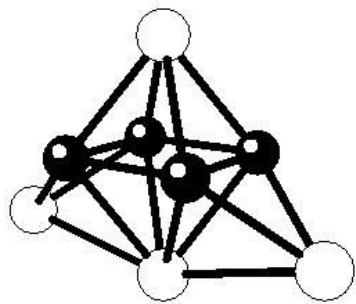
(a)

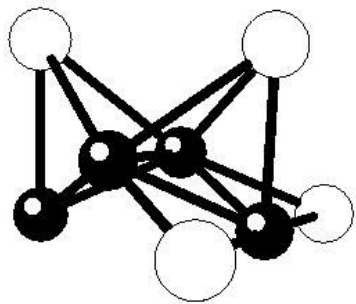
(b)

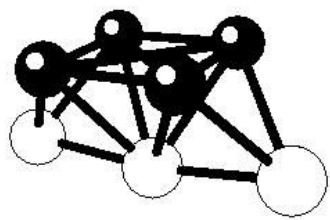
(c)

Fig. 1

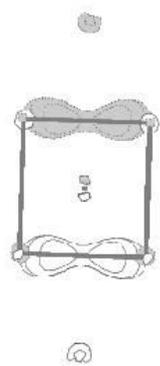 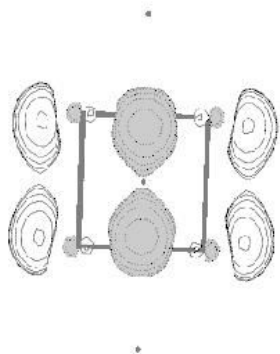 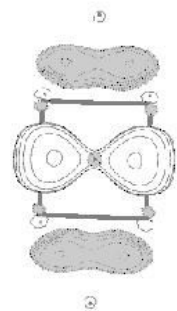

HOMO            HOMO-1            HOMO-2

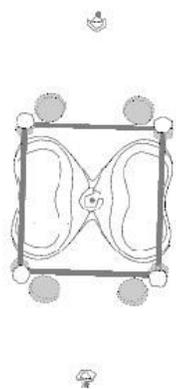 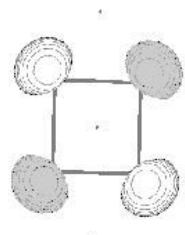

HOMO-3          HOMO-4

Fig. 2

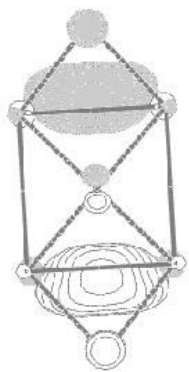
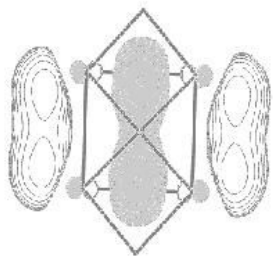
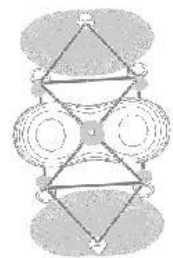

HOMO  HOMO-1  HOMO-2

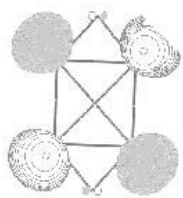
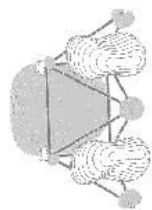

HOMO-3  HOMO-4

Fig. 3

# Metallo-Anti-aromatic $Al_4Na_4$ and $Al_4Na_3^-$: A Theoretical Investigation

Sharan Shetty, Dilip G. Kanhere, Sourav Pal*

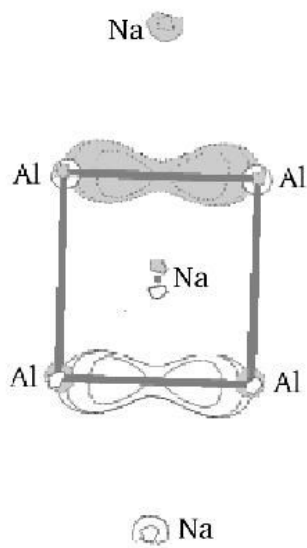

Present theoretical investigation reveals that the neutral $Al_4Na_4$ and anion $Al_4Na_3^-$ clusters satisfy the criteria for anti-aromaticity and are metallo-anti-aromatic compounds.